\documentclass[final,5p,times,twocolumn]{elsarticle}

\usepackage{graphicx}
\usepackage{amssymb}

\usepackage{ulem} 
\usepackage[usenames]{color}
\newcommand{\be}{\begin{eqnarray}}
\newcommand{\ee}{\end{eqnarray}}

\begin{document}

\begin{frontmatter}

\author[juli,ias]{S.~Krewald}
\ead{s.krewald@fz-juelich.de}
\author[bo]{E.~Epelbaum} 
\author[juli,ias,bonn]{U.-G.~Mei\ss ner} 
\author[juli]{P.~Saviankou}
\address[juli]{Institut f\"ur Kernphysik and J\"ulich Center for Hadron Physics, 
Forschungszentrum J\"ulich, D-52425 J\"ulich,Germany}
\address[ias]{Institute for Advanced Simulation,
Forschungszentrum J\"ulich, D-52425 J\"ulich,Germany}
\address[bo]{Institut f\"ur Theoretische Physik II, Ruhr-Universit\"at-Bochum,
D-44780 Bochum, Germany}
\address[bonn]{Helmholtz-Institut f\"ur Strahlen- und Kernphysik (Theorie) and Bethe Center for Theoretical Physics, 
Universit\"at Bonn, D-53115 Bonn, Germany
}
\title{Saturation of nuclear matter in effective field theory}

\begin{abstract}
The two- and three-nucleon interaction derived in chiral effective field theory
at next-to-next-to-leading order 
is used to obtain the binding energy of nuclear matter. Saturation is found at a binding energy 
per particle $E/A = -16.2\pm0.3$~MeV and a Fermi momentum $k_F = 1.30 \pm 0.03$~fm$^{-1}$, where the
uncertainty is due to the cut-off dependence of the two-nucleon
interaction. The sensitivity of these values to the three-nucleon force is
also studied.
\end{abstract}


\end{frontmatter}

\section{Introduction}

Within the last two decades, effective field theory (EFT) has  provided new
methods for nuclear structure investigations. Both two- and three-nucleon
interactions based on the most general chiral effective pion-nucleon Lagrangian
have been developed utilizing Weinberg's power  counting scheme.
The phase shifts of the nucleon-nucleon interaction below the pion production threshold
are reproduced with a precision that is comparable with the one achieved by modern 
phenomenological two-nucleon potentials, when pushing the expansion to the
order N$^3$LO,\footnote{Throughout, we use the abbreviation LO for leading
  order, NLO for next-to-leading order and so on.}
see Refs.\cite{Entem:2001cg,Epelbaum:2004fk}. These forces together with
three-nucleon forces at N$^2$LO have been applied to a cornucopia of reactions.
In particular, a good description of most of the low-energy nucleon-deuteron 
scattering observables at order N$^2$LO has been reported. 
The binding energies of $^3$He and $^4$He are also correctly reproduced 
once the leading three-nucleon force is included.
Within the framework of the no-core shell model, the chiral forces
have been used to study nuclei with $A = 7 , 10 \ldots 13$ systems, in
particular the sensitivity to the three-nucleon force. For details, 
see \cite{Nogga:2005hp,Navratil:2007we}. 
Recent progress along these lines is reviewed in Ref.~\cite{Epelbaum:2008ga}.

Given these successes, the question arises whether chiral effective 
field theory can contribute to the solution of the nuclear matter problem. First attempts in
this direction have been done by the Munich group, see e.g. 
Refs.~\cite{Kaiser:2001jx,Kaiser:2006tu} (and references 
therein)
which concentrates on the development of a chiral effective interaction
to be used exclusively in the nuclear medium without link to the two-nucleon interaction
in the vacuum.
In Ref.\cite{Bogner:2005sn},
nuclear matter has been studied with a low-momentum two-nucleon force $V_{\rm low-k}$ derived
from the Argonne $v_{18}$ potential via renormalization group techniques.
It was found that the low-momentum interactions can be used in second-order perturbation theory
to describe nuclear matter provided Pauli blocking effects due to the medium
are properly included. This approach was recently extended to the $V_{\rm
  low-k}$ versions of the chiral N$^3$LO potentials~\cite{Bogner:2009un}.
Note, however, that while the chiral
three-body force at N$^2$LO has been taken into account, its running down to
low momenta has not been considered.

 In the present communication, we apply the  N$^2$LO version
 of the chiral nucleon-nucleon interaction to nuclear matter.
 We  use chiral nuclear  effective field theory because it gives a systematic
connection between the two- and many-nucleon forces in each order of the
chiral expansion, as reviewed in  Ref.~\cite{Epelbaum:2008ga}.
We also compare with two other recent approaches to nuclear matter 
based on LO versions of the
two-nucleon interactions~\cite{Lacour:2010AP,Machleidt:2010PR}.

\section{The method}

We employ the two-nucleon interaction derived by Epelbaum, Gl\"ockle, and Mei{\ss}ner
using effective field theory (EFT)~\cite{Epelbaum:2004fk}. 
Two regulators are introduced in Ref.~\cite{Epelbaum:2004fk}, 
a cut-off $\tilde{\Lambda}$ for the  spectral function
representation  of the two-pion exchange diagrams and 
a cut-off $\Lambda$ required for the non-perturbative renormalization of the
Lippmann-Schwinger equation.
As in  Ref.\cite{Epelbaum:1998ka},
 the pion-nucleon vertices are systematically taken from the chiral
expansion of the pion-nucleon interaction which provide the direct link
to the spontaneously broken chiral symmetry of QCD.
In the present calculations, we adopt the values (in units 
of GeV$^{-1}$)
\begin{eqnarray}\label{eq:11}
c_1=-0.81,~ c_2=3.28,~ c_3=-3.4,~ c_4=3.4~,
\end{eqnarray}
 employed in Ref.\cite{Epelbaum:2004fk} which enter the
long-range part of the two- and three-nucleon force.
The two-nucleon interaction is generated by eliminating  the pion-two-nucleon space 
via the time-honored projection method of
  Fukuda, Sawada, Taketani, and  Okubo~\cite{FST,Okubo}.
A discussion of the sensitivity of the observables  to  the cutoffs 
can be found in Ref.~\cite{Epelbaum:2004fk} where, in particular, the
error bands for the phase shifts are shown.

The phenomenological two-nucleon interactions of the 1950's
 such as e.g.~the Reid potential~\cite{reid}
could not be used in perturbative approaches to nuclear matter because a strong repulsion at
short distances between the interacting nucleons was required to reproduce the empirical  
S-wave phase shifts. Brueckner proposed an effective two-nucleon interaction, commonly called the
G-matrix, which allows to perform perturbative calculations of nuclear matter
\cite{Brueckner58,day_67}. 
Nuclear matter calculations within the Brueckner theory
based on phase shift equivalent potentials 
yield  the well-known Coester band~\cite{Coester:1970} which does not pass through the 
empirical saturation energy $E/A=-16\,$MeV at the Fermi momentum $k_F = 1.36 \,{\rm fm}^{-1}$. 
Theoretical efforts to include the effect of three-nucleon correlations  
show that three-nucleon correlations by themselves do not suffice to solve the nuclear matter
problem, see e.g. Refs.\cite{Raj_Bethe67,Bethe71,DAY81,song:1998,sartor:2006}.
The necessity to incorporate  three nucleon interactions in a theory of nuclear matter 
was realized subsequently, see Ref.\cite{schiavilla}, but progress was made difficult by the
limited experimental information about three-nucleon interactions.

In  1971, Baker criticized  Brueckner theory.
In order to obtain a well-defined analytical structure of the energy function, he
rearranged the energy series in powers of the R-matrix which is the solution of the
Lippmann-Schwinger integral equation for nucleon-nucleon scattering 
incorporating  Pauli blocking of the intermediate states due to the nuclear
 medium ~\cite{Baker76}.
While Baker has illustrated his formalism using a two-nucleon interaction
consisting of a short-range hard core and a medium-range constant 
attractive potential,
 we want to operate with the interactions derived in effective field theory.

Our calculations proceed in three steps:

(i)~ The R-matrix is obtained by solving the Lippmann-Schwinger equation
in a partial wave representation, using the matrix inversion method of
Haftel and Tabakin~\cite{HT1970}: 
\begin{eqnarray}\label{eq:1}
&&R(K,q_f,q_i)^{\alpha}_{L_f,L_i} = V(q_f,q_i)^{\alpha}_{L_f,L_i} 
         - \frac {2}{\pi} \sum_{L'} \int dq' q'^2 
\nonumber \\ 
&& \times V(q_f,q')^{\alpha}_{L_f,L'} Q(K,q') g(q',q_i)
R(K,q',q_i)^{\alpha}_{L',L_i}. 
\nonumber \\
\end{eqnarray}
The angular momentum $J$, the spin $S$, and the isospin $I$
  are combined into the index $\alpha$,
while the orbital angular momenta 
 in the initial and final state are denoted by 
$L_i$, and $L_f$, respectively. The magnitudes of the relative momenta
of the interacting nucleons are called
 $q_i$, $q'$, and $q_f$.
The angle-averaged Pauli operator is denoted  $ Q(K,q')$, where $K$ stands for
the center-of-mass momentum of the interacting particles.
The two-nucleon propagator $g(q',q)$ is regularized by a finite imaginary contribution
$\Gamma$
\begin{equation}
 g(q',q) =  \frac {q'^2 -q^2} {( q'^2 - q^2 )^2 + \Gamma^2}.
\end{equation}
The calculations are performed for three different values of $\Gamma$ and
to obtain the binding energy of nuclear matter, an extrapolation to $\Gamma = 0$
is performed.

(ii) ~The contribution of the two-body interaction to the binding energy of
nuclear matter will be denoted as $(E/A)_2$ and is given by:
\begin{eqnarray}\label{eq:2}
(E/A)_2 &=& \frac {4}{\pi m_N} \sum_{\alpha, L} ( 2J+1 ) (2T+1) k_F^3
\nonumber \\ 
&\times&\int_0^1 dx x^2 w(x) 
R(K(x) k_F ,x k_F, x k_F)^{\alpha}_{L,L}\, , 
\end{eqnarray}
where an average momentum 
\begin{equation}
K(x)^2 = \frac {3}{5} ( 1 - x ) \left( 1 + \frac {x^2}{3 ( 2 + x )} \right)
\end{equation}
and a weight factor 
\begin{equation}
w(x) = 1 - \frac{3}{2} x + \frac{1}{2} x^3  
\end{equation}
are used and $m_N$ is the nucleon mass.

(iii) ~The contribution of the three-nucleon forces to the binding energy
reads:
\begin{eqnarray}\label{eq:3}
&&(E/A)_3 = \frac {g_A^2}{(2 \pi f_{\pi})^4 } \frac {36}{ \pi k_F^3 } 
\int_0^{k_F} dP_{23} P_{23}^2
\nonumber \\ 
&&\times \left(\int_0^{k_F-P_{23}} dp p^2 \int_{-1}^{+1} d\zeta   +
  \int_{k_F-P_{23} }^{\sqrt{k_F^2 - P_{23}^2} } dp p^2  
   \int_{-\zeta_{\rm max} }^{ \zeta_{\rm max} } d\zeta \right)
\nonumber \\ 
&&\times \left(\int_0^{k_F-P_{23}} dq' q'^2 \int_{-1}^{+1} d\eta   +
  \int_{k_F-P_{23} }^{k_F + P_{23} } dq' q'^2  
   \int_{-1 }^{ \eta_{\rm max} } d\eta \right) 
\nonumber \\ 
&&\times \int_0^{2\pi}
d\phi T_{\Lambda}(p,q')^2
\left( c_1 T_1 + c_3 T_3 + c_4 T_4 + c_D T_D + c_E T_E \right),
\nonumber\\
\end{eqnarray}
using
\begin{eqnarray}\label{eq:4}
T_1 & = & -\frac {k_{12}^2} {k_{12}^2+M_{\pi}^2}
\frac {M_{\pi}^2} {k_{23}^2+M_{\pi}^2}
 - 2 \frac {k_{12}^2 M_{\pi}^2} {(k_{12}^2+M_{\pi}^2)^2} 
\nonumber \\ 
T_3 & = & \frac {(\vec{k}_{12}\cdot \vec{k}_{23})^2 } 
             { 2(k_{12}^2+M_{\pi}^2)(k_{23}^2+M_{\pi}^2) }
 - \frac {k_{12}^4 } {(k_{12}^2+M_{\pi}^2)^2} 
\nonumber \\ 
T_4 & = & \frac {(\vec{k}_{12}\times \vec{k}_{23})  \cdot
                 (\vec{k}_{12}\times \vec{k}_{23}) }
             { 2(k_{12}^2+M_{\pi}^2)(k_{23}^2+M_{\pi}^2) } 
\nonumber \\ 
T_D & = & \frac {p^2}{g_A(4p^2+M_{\pi}^2)} 
\nonumber \\ 
T_E & = & -\frac {1}{g_A^2}\,,
\end{eqnarray}
with $M_\pi$ denoting the pion mass and $g_A$ the
nucleon axial-vector coupling constant. 
The momenta 
$\vec{k}_{ij} =  \vec{k}_{i} - \vec{k}_{j}$ denote the relative momenta
of the
 single nucleon momenta $\vec{k}_1,\vec k_2, \vec k_3$. 
 In terms of the Jacobi vectors used in Eq.(\ref{eq:3}), the single
nucleon momenta read
\begin{eqnarray}\label{eq:5}
\vec{k}_1 & = & \vec{P}_{23} + \vec{q}\,' ,
\nonumber \\ 
\vec{k}_2 & = & \vec{P}_{23} + \vec{p},
\nonumber \\ 
\vec{k}_3 & = & \vec{P}_{23} - \vec{p},
\end{eqnarray}
which implies  $\vec{P}_{23} = (\vec{k}_2+ \vec{k}_3)/2,$ 
$\vec{p} = (\vec{k}_2 -\vec{k}_3)/2,$ 
$\vec{q}\,' = \vec{k}_1 -\vec{P}_{23}.$ 
The quantity $\zeta = \cos(\theta_p)$ stands for the cosinus of the angle
$\theta_p$ between the vectors $\vec{P}_{23}$ and $\vec{p}$. The position of the
vector $\vec{q'}$ relative to $\vec{P}_{23}$ is characterized in Eq.~(\ref{eq:3})
by the two angles $\theta_{q'}$ and $\phi$. We abbreviate 
$\eta = \cos(\theta_{q'})$,
$$\eta_{\rm max}=({k_F^2-P_{23}^2-q'^2})/({2P_{23}q'}),$$
and 
$$\zeta_{\rm max}=({k_F^2-P_{23}^2-p^2})/({2P_{23}p}).$$
%
We use the same regulator function for the three-nucleon
interaction as in  Ref.~\cite{Epelbaum:2004fk}:
\begin{eqnarray}\label{eq:6}
T_{\Lambda}(p,q') = \exp\left(-\left(\frac {p}{\Lambda}\right)^2
 -\frac{3}{4} \left(\frac {2 q'}{3 \Lambda}\right)^2 \right).
\end{eqnarray}

The pion-nucleon low-energy constants $c_1, c_3$, and $c_4$ shown in Eq.~(\ref{eq:11})
are taken from 
Ref.~\cite{Epelbaum:2004fk}. The short-range part of the  three-nucleon interaction
is characterized by the low-energy constants $c_D$ and $c_E$. 
In principle, these constants can be obtained from the analysis of few-nucleon
systems. 
The values, however, are strongly scheme and cutoff dependent. In addition,
they have not yet been determined with the same precision as e.g. the 
pion-nucleon LECs $c_i$  \cite{Epelbaum:2008ga}.
For definiteness, we employ  the values $c_E = -c_D = 0.2$ in the present calculations
 which are of natural size and
leave an improved determination of those constants for later work.
We will, however, explore the sensitivity of the binding energy per particle
to the values of $c_E$ and $c_D$ varied within a natural range.

Finally, the energy per particle of nuclear matter is the sum of the kinetic energy
and the two contributions of Eqs.~(\ref{eq:2}) and (\ref{eq:3}).
\section{ Results at next-to-next-to-leading order}

\begin{figure}[t]
\begin{center}
   \includegraphics[height=0.5\textwidth,angle=-90]{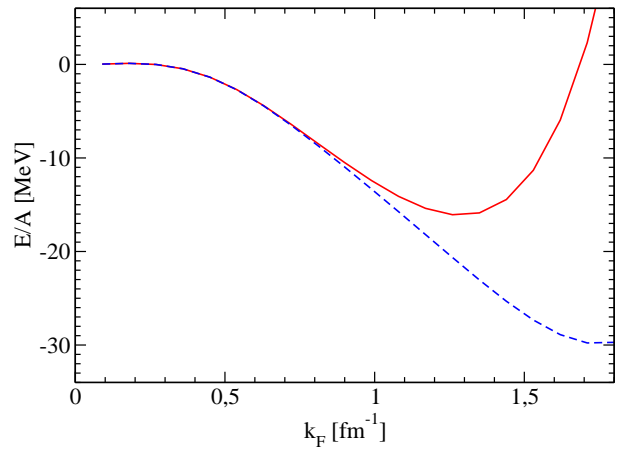} 
\caption{
Binding energy per particle of nuclear matter as a function of the
Fermi momentum $k_F$ for the NNLO potential using  
a Lippman-Schwinger cut off $\Lambda = 550$~MeV and a spectral function
cut off $\tilde{\Lambda} = 600$~MeV (solid(red) line). 
The binding energies obtained in the absence of three-nucleon interactions
are shown by the dashed (blue) line.
 }
\label{fig:1}  
\end{center}  
\end{figure}

At leading order and next-to-leading order, only two-body forces appear.
The description  of nuclear matter is comparable to what has been
found using other two-nucleon interactions which reproduce the two-nucleon phase shifts.
The binding energy of nuclear matter derived from effective two-nucleon forces in order NLO
saturates well above $k_F = 1.5 \, {\rm fm}^{-1}$ and will not be discussed in detail because it is
quite similar to the result one obtains in N$^2$LO after switching off the three-nucleon
interactions, see the dashed (blue) line in  Fig.~\ref{fig:1}.\

Recently, two calculations of nuclear matter based on interactions derived from
Effective Field Theory have been published.
Ref.~\cite{Lacour:2010AP} employs a novel two-nucleon interaction obtained
in LO to determine the binding energy  of nuclear matter and show that
saturation of nuclear matter can be obtained qualitatively already at LO. 
Their scheme is different from ours as is establishes an in-medium power
counting while  our approach employs the amplitudes as given by the
vacuum power counting. The  Fermi momenta at the saturation point range from 
0.47~fm$^{-1}$ to 0.93~fm$^{-1}$ which is below the empirical one, but
after inclusion of a phenomenological correction,  which is formally of 
higher order, the empirical saturation point can be reproduced. In particular, the
work of Ref.~\cite{Lacour:2010AP} justifies the earlier work of the
Munich group \cite{Kaiser:2001jx}, in which multi-nucleon interactions
were not considered.

Ref.~\cite{Machleidt:2010PR} 
 finds underbinding of nuclear matter at LO,
 achieving only $-10$~MeV binding energy at
a Fermi momentum of 1.8~fm$^{-1}$.
Ref.~\cite{Machleidt:2010PR} employs the standard
choice of the mean field of nuclear matter ($U(k)=0~{\rm for}~ k>k_F$).
Note, however, that Jeukenne et al. ~\cite{Jeukenne76}
a long time ago have criticized this standard choice
 because it violates the
analytical structure of the self-energy of the nucleon, and  recommended to
use a mean field continuous at  the Fermi surface.
Refs.~\cite{song:1998,sartor:2006}
show that Brueckner calculations
 employing the standard choice of the mean field underestimate the binding
energy of nuclear matter as compared to calculations employing a continuous mean field.

The binding energy of nuclear matter  computed in the absence of
 three-nucleon forces, using only the R-matrix,
is shown by the  dashed (blue) line in  Fig.~\ref{fig:1}.
 There is a saturation point
 near $ k_F = 1.75 \, {\rm fm}^{-1}$. Assuming that this result is within the
 validity range of  EFT, this confirms the finding of
Ref.~\cite{Lacour:2010AP} that EFT employing a
two-nucleon interaction which fits the phase shifts  produces 
saturation of nuclear matter. For a quantitative description
of the empirical saturation point, one has to go beyond LO, however.

In the low-density limit, we differ from the LO predictions of 
Ref.~\cite{Lacour:2010AP} which finds a rapid increase of the 
 binding energy for small densities. 
The present approach produces positive energies for nuclear matter
at small densities,
implying that nuclear matter is not the energetically most 
favorable state. The critical Fermi momentum for a transition to a deuteron gas
can be estimated using the binding energy of the deuteron, $E_d = -2.225\,$MeV,
one finds $(k_F)_{\rm critical} = 0.51 ~{\rm fm}^{-1}$.

In the counting scheme of (vacuum) EFT, three-nucleon interactions
appear for the first time at order N$^2$LO (if an energy-independent scheme
to calculate the potential is utilized).  For our central choice 
  $c_E = -c_D = 0.2$, 
the saturation curve obtained (solid red line) has a saturation point
at the Fermi momentum $k_F=1.29$~fm$^{-1}$, where the maximum binding energy of
nuclear matter is $E/A= -16.11\,$MeV, which is quite an amazing result.  
 We do not show binding energies of
nuclear matter above $ k_F = 1.8~{\rm fm}^{-1}$ because the corresponding kinetic
energies of incident nucleons in nucleon-nucleon scattering would be above
the pion production threshold.

\begin{figure}[t]
\begin{center}
\resizebox{0.45\textwidth}{!}{ 
  \includegraphics[angle=-90]{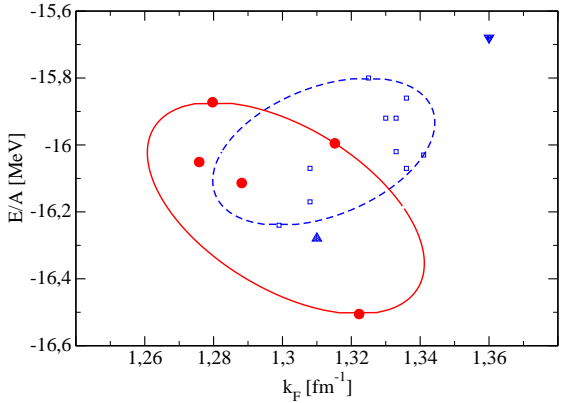} 
}

\vspace{3mm}

\caption{
Saturation points of nuclear matter.
Downward triangle (blue): Bethe's estimate\cite{Bethe71},
upward  triangle (blue): M\"oller, Myers, Swiatecki, and Treiner \cite{Myers:1988},
rectangle (blue): microscopic mean-field approaches\cite{Bender2003},
circles (red): present approach.
 }
\label{fig:2}  
\end{center}  
\end{figure}
At order N$^2$LO, Ref.~\cite{Epelbaum:2004fk} provides 
two-nucleon interactions for five different combinations of regulators.
In Fig.~\ref{fig:2}, we show the saturation points of nuclear matter obtained
in the present formalism for each of these interactions  by the filled (red) circles.
The binding energies per particle scatter between $-15.8$~MeV and $-16.6$~MeV,
while the Fermi momenta are contained in the interval ranging from 
  $ k_F = 1.27~{\rm fm}^{-1}$ to 
  $ k_F = 1.33~{\rm fm}^{-1}$. This is quite an encouraging result, given
that we have not performed any fine-tuning on the input parameters.
For comparison, let us discuss some earlier determination of these
quantities.
Bethe employed the semi-empirical mass formula for finite nuclei by Myers and
Swiatecki~\cite{Myers:1966} 
to obtain the binding energy of nuclear matter  
and estimated the Fermi momentum using the central density
of finite nuclei and the argument that the surface tension is balanced by the
Coulomb pressure~\cite{Bethe71} (downward (blue) triangle). A more detailed treatment of the
Coulomb force is made by microscopic calculations employing empirical interactions
which need approximately ten parameters to reproduce both the binding energies 
and the charge distributions of finite nuclei. The review by Bender, Heenen,
and Reinhard~\cite{Bender2003}
discusses the merits of the various types of phenomenological 
interactions which have been adjusted to reproduce the binding energies and radii
of finite nuclei as accurately as possible.
The corresponding predictions for the saturation of nuclear matter are
summarized by the blue squares.
A (blue) dashed ellipse has been added to guide the eye.
The finite-range droplet model of M\"oller, Myers, Swiatecki, and Treiner
generalizes the liquid drop model by including shell corrections~\cite{Myers:1988}.
The corresponding saturation point of nuclear matter,
$E/A = -16.279$~MeV and k$_F$ = 1.31~fm$^{-1}$, is indicated by the upward
(blue) triangle.
By inspection one finds that the saturation points obtained by the 
present calculation based on the N$^2$LO interaction (circles) overlap with the
empirical ones. There is a dependence on the regulators 
   $\Lambda$  and 
  $\tilde{\Lambda} $ which leads to an uncertainty of the
binding energy per particle of approximately 0.6~MeV, while the Fermi momenta
scatter within an interval of $0.06~{\rm fm}^{-1}$. The magnitude of this uncertainty is
comparable to the uncertainty due to the different types of empirical interactions,
however.

\begin{figure}[t]
\begin{center}
\resizebox{0.44\textwidth}{!}{ 
  \includegraphics[angle=0]{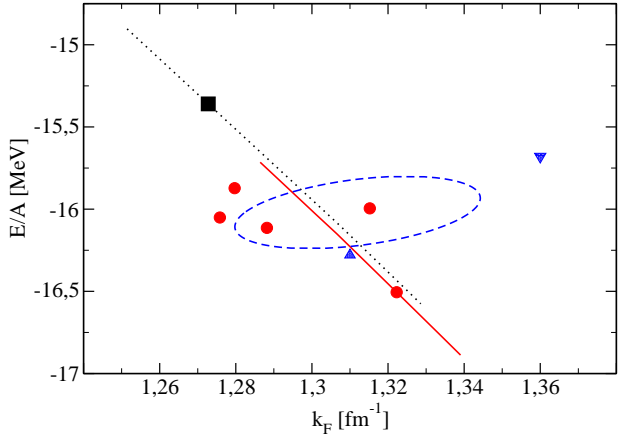} 
}

\vspace{3mm}

\caption{
The dependence of the saturation point on the three-nucleon interaction.
The dashed (black) line shows the saturation points when $c_D$ is changed from
$-1.1$ to $+0.5$ for $c_E = 0.0$, the (black) square denotes the point corresponding to
$c_D = 0$ and $c_E = 0$. The solid (red) curve keeps $c_E = 0.2$ and varies
$c_D$ between $-0.5$ and $+0.5$. The (red) circles  show the saturation points
obtained in the present approach.
Downward triangle(blue): Bethe's estimate~\cite{Bethe71},
upward  triangle(blue): M\"oller, Myers, Swiatecki, and Treiner \cite{Myers:1988},
ellipse(blue): microscopic mean-field approaches~\cite{Bender2003}.
 }
\label{fig:3}  
\end{center}  
\end{figure}

There is another uncertainty due to the three-nucleon interaction. 
In Fig.~\ref{fig:3},  we compare the saturation point of nuclear matter obtained
using only the longest range part of the three-nucleon interaction 
(square, obtained with $c_D=0$ and $c_E=0$) with the result of the full
calculation. The corresponding binding energy, E/A = $-15.36\,$MeV, and the
Fermi momentum $k_F = 1.27$~fm$^{-1}$ are in the vicinity of the empirical saturation
area. In order to show the sensitivity to the short range contributions, we have varied
the value of $c_D$ in the range from $-1.1$ to $+ 0.5$ for $c_E=0$ and $-0.5$ to $+0.5$ for
$ c_E=0.2$. These ranges are consistent with determinations of the LECs from
few-nucleon reactions, see~Ref.~\cite{Epelbaum:2008ga}.
We find that the empirical saturation area can be approached with relatively
small values of both $c_E$ and $c_D$ which suggests that Effective Field Theory applied
to nuclear matter converges reasonably fast. The uncertainty inherent in the two-nucleon
interaction at order N$^2$LO is comparable to what is obtained in a systematic comparison
of high quality fits of more phenomenological interactions, see Ref.~\cite{Bender2003}.
Therefore a detailed fine-tuning of the low energy constants $c_D$ and $c_E$ is not
necessary.

The role of the delta-isobar in the description of nuclear matter based
on chiral dynamics was stressed in Ref.~\cite{Fritsch:2004nx}. 
Including the $\Delta_{33}$ resonance explicitly in an effective field theory 
treatment of pion-nucleon
scattering, one finds the following values of the 
low energy constants $c_1, c_2, c_3$, and $c_4$ of the pion nucleon Lagrangian,
see Ref.~\cite{Krebs:2007rh},
$c_1=-0.57,~ c_2=-0.25,~ c_3=-0.79,~ c_4=1.33$,
in units of GeV$^{-1}$.
Comparing with Eq.(\ref{eq:11}), one notices a reduction of the magnitudes of
those constants. It is an interesting question for future work to find out whether
an explicit treatment of the $\Delta(1232)$ may help to minimize the remaining 
uncertainties of the effective field theory approach to nuclear matter.

The compression modulus of nuclear matter,
\begin{equation}
K=k_F^2\frac{\partial^2E/A}{\partial k_F^2}~,
\end{equation}
is found to be $K = 197$~MeV, with a spread 
$K_{\rm max} - K_{\rm min} = 37$~MeV.
By analyzing the mean excitation energy of the
isoscalar giant resonance in heavy nuclei with a set of 
effective interactions defined in the nuclear medium, Blaizot found  values
$ K = 210 \pm 30$~MeV~\cite{Blaizot80}.
The interactions adjusted directly to the saturation properties of finite nuclei
produces compression moduli of nuclear matter ranging from $K = 172$~MeV to
$K = 356$~MeV~\cite{Bender2003}. 
In order to characterize the saturation curve above the saturating momentum, we
suggest to complement the compression modulus of nuclear matter by the 
skewness parameter 
 $S=k_F^3 ({\partial^3E/A})/({\partial k_F^3})$. 
The present calculation finds  $S = 914 \pm 217$~ MeV.  It would
be interesting to see, what values for $S$ are found e.g. in the Munich approach
or in more phenomenological investigations.

\section{Conclusions}
We have shown that a  many-body theory of nuclear matter employing 
the two- and three-nucleon interactions derived from effective field theory at N$^2$LO 
produces saturation of nuclear matter at binding energies $E/A =  -16.2 \pm 0.4$~MeV and
Fermi momenta $k_F = 1.30 \pm 0.03$~fm$^{-1}$ close to the empirical value obtained from the 
finite range liquid drop model~\cite{Myers:1988}. 
{The uncertainties in the binding energy and saturating Fermi momentum of nuclear matter
shown above are due to the dependence of the two-nucleon interaction
on the regulators $\Lambda$ and $\tilde{\Lambda}$.
The uncertainties due to the low-energy constants $c_E$ and $c_D$
have not been studied in detail. Changing $c_E$ and $c_D$ in a range of natural values of 
either sign, one finds a line that passes through the region of saturation points produced
by phenomenological approaches that provide the best fits to the binding energies and radii of
finite nuclei, see the compilation in Ref.~\cite{Bender2003} 
and  Fig.~\ref{fig:3}.} 
The nuclear compressibility is in the range from  179 to 216~MeV.
Nuclear matter is found to be instable for Fermi momenta below 0.51 fm$^{-1}$,
where a transition to a deuteron gas is expected.\\

We thank A. Nogga and V. Tselyaev for helpful comments.
We thank the DFG for partial support by
the grant GZ:432RUS113/904/0-1. This work was further supported in 
parts by funds provided from the 
Helmholtz Association to the young investigator group  
``Few-Nucleon Systems in Chiral Effective Field Theory'' (grant  VH-NG-222)
and the Virtual Institute ``Spin and strong QCD'' (grant VH-VI-231). 
This work was further supported by the DFG (SFB/TR 16 ``Subnuclear Structure
of Matter'') and  the European Community-Research Infrastructure
Integrating Activity ``Study of Strongly Interacting Matter''
(acronym HadronPhysics2, Grant Agreement n. 227431)
under the Seventh Framework Programme of the EU, and by BMBF (grant 
06BN9006).

\end{document}